# Around which stars can TESS detect Earth-like planets? The Revised TESS Habitable Zone Catalog

L. Kaltenegger[1], J.Pepper[2], P. M. Christodoulou[2], K. Stassun[3], S. Quinn[4], C. Burke[5]
[1] Carl Sagan Institute, Cornell University, Space Science Institute 312, 14850 Ithaca, NY, USA,
lkaltenegger@astro.cornell.edu, Tel: +1-607-255-3507
[2] Lehigh University, Physics Department, Bethlehem, PA 18015, USA
[3] Vanderbilt University, Physics & Astronomy Department, Nashville, TN 37235, USA
[4] Center for Astrophysics | Harvard & Smithsonian, 60 Garden street, Cambridge, MA, USA
[5] MIT Kavli Institute for Astrophysics and Space Research, 77 Massachusetts Avenue, Cambridge, MA 02139, USA

## Abstract

In the search for life in the cosmos, NASA's Transiting Exoplanet Survey Satellite (TESS) mission has already monitored about 74% of the sky for transiting extrasolar planets, including potentially habitable worlds. However, TESS only observed a fraction of the stars long enough to be able to find planets like Earth.

We use the primary mission data – the first two years of observations – and identify 4,239 stars within 210pc that TESS observed long enough to see 3 transits of an exoplanet that receives similar irradiation to Earth: 738 of these stars are located within 30pc. We provide reliable stellar parameters from the TESS Input Catalog that incorporates Gaia DR2 and also calculate the transit depth and radial velocity semi-amplitude for an Earth-analog planet.

Of the 4,239stars in the Revised TESS HZ Catalog, 9 are known exoplanet hosts – GJ1061, GJ1132, GJ3512, GJ685, Kepler-42, LHS1815, L98-59, RRCae, TOI700 – around which TESS could identify additional Earth-like planetary companions. 37 additional stars host yet unconfirmed TESS Objects of Interest: three of these orbit in the habitable zone – TOI203, TOI715, and TOI2298.

For a subset of 614 of the 4,239stars, TESS has observed the star long enough to be able to observe planets throughout the full temperate, habitable zone out to the equivalent of Mars' orbit.

Thus, the Revised TESS Habitable Zone Catalog provides a tool for observers to prioritize stars for follow-up observation to discover life in the cosmos. These stars are the best path towards the discovery of habitable planets using the TESS mission data.

## 1. Introduction

Several thousand exoplanets have already been discovered, among them the first dozen terrestrial planets orbiting in the temperate so-called Habitable Zone (HZ) (see e.g., Kasting et al. 1993) of their stars, which could provide habitable conditions (e.g. Kane et al. 2016, Johns et al. 2018; Berger et al. 2018). How many stars has NASA's Transiting Exoplanet Survey Satellite (TESS) mission already observed long enough to identify transiting exoplanets which receive similar irradiation to Earth? And for how many stars did TESS already collect enough data to observe transiting exoplanets throughout the whole Habitable Zone?

NASA's Transiting Exoplanet Survey Satellite (TESS) mission (Ricker et al. 2016) has already searched about 74% of the sky in its two-year primary mission for transiting extrasolar planets, monitoring several hundred thousand stars. TESS's primary mission is adding exoplanets (e.g., Kostov et al. 2019, Vanderspek et al 2019, Luque et al 2019, Huang et al 2018) and more than two thousand planetary candidates – as TESS Objects of Interest (TOIs; see Guerrero et al. 2020) – to the list, including the first potentially habitable world discovered by TESS, TOI 700d (Gilbert et al. 2020, Rodriguez et al. 2020).

TESS observes the sky in individual sectors for about 27 days each, with sectors overlapping toward the ecliptic pole. This strategy results in stars closer to the ecliptic being observed for about 27 days, and stars close to the ecliptic poles for nearly a year. The region with the longest observation time overlaps with JWST's continuous viewing zone.

The TESS Habitable Zone Star Catalog (Kaltenegger et al. 2019a) initially derived a list of stars where TESS was expected to be able to detect transiting planets with Earth-analog irradiation based on expected observation times (see discussion). Here we update this star list by using the real observations from the TESS prime mission: we identify 4,239 stars that TESS has observed long enough to see at least three transits of planets receiving the equivalent irradiation to modern Earth (data available online and on *filtergraph.com/tesshzstars*).

Nine stars in the Revised TESS HZ Catalog are known exoplanet hosts around which TESS could identify additional planetary companions with the same irradiation as Earth. 37 additional stars host yet unconfirmed TOIs, with three of these candidates orbiting in the HZ of their star, which we discuss in detail.

We also identify the subset of 614 stars for which TESS could have observed three transits for exoplanets out to the outer edge of the HZ. This subset of stars can give insights into the population of transiting planets throughout the HZ.

## 2. Methods

### 2.1 Habitable Zone Model

When searching for life in the universe, the HZ is a concept that helps guide observations: It identifies the orbital distance region around one or multiple stars where liquid water could be stable on a rocky planet's surface similar to Earth (e.g., Kasting et al. 1993, Stevenson 1999, Pierrehumbert & Gaidos 2011, Kaltenegger & Haghighipour 2013, Kane & Hinkel 2013, Kopparapu et al. 2013, 2014, Ramirez & Kaltenegger 2017, 2018). We use liquid surface water because it has not yet been demonstrated that subsurface biospheres – e.g. under an ice layer on a frozen planet – can modify a planet's atmosphere so that its existence can be detected remotely.

The width and orbital distance of the HZ depend, to first approximation, on two main parameters: incident stellar flux and planetary atmospheric composition (see e.g., review by Kaltenegger 2017). Here we use the empirical HZ limits for an Earth-analog rocky planet with 1 Earth mass and 1 Earth radius a mostly $N_2$-$H_2O$-$CO_2$ atmosphere (see e.g., Kasting et al. 1993), to model the range of relevant orbital distances. The empirical HZ is based on stellar irradiation received by a young Venus and a young Mars when there is no more evidence for liquid surface water as originally defined using a 1D climate model by Kasting et al. 1993. Note that the inner limit based on data from Venus is not well known because of the lack of reliable geological surface history beyond 1 billion years ago.

We use a 4$^{th}$ order polynomial fit for the empirical HZ limits around main sequence stars with surface temperatures based on values derived from models by (Kasting et al 1993, Kopparapu et al. 2013, 2014 and an extension of that work to 10,000K by Ramirez & Kaltenegger 2016):

$$S_{\text{eff}} = S_{\text{Sun}} + aT + bT^2 + cT^3 + dT^4 \quad (1)$$

where T* = ($T_{\text{eff}}$ – 5780) and $S_{\text{sun}}$ is the stellar incident values at the HZ boundaries in our solar system. Table 1 provides the values to estimate the stellar irradiation at the boundaries of the HZ for different host stars using equation 1. The orbital distance of the HZ boundaries can then be calculated from $S_{\text{eff}}$ using equation 2:

$$a = \sqrt{\frac{L/L_{\text{Sun}}}{S_{\text{eff}}}} \quad (2)$$

where $L/L_{sun}$ is the stellar luminosity in solar units and $a$ is the orbital distance in AU. The corresponding period can be derived using equation 3. If the orbital distance of a detected planet is larger than $a_{RV}$ and smaller than $a_{EM}$, then the detected planet is orbiting its host star in the empirical HZ. The corresponding period of the orbit at the boundaries of the empirical HZ can be expressed using equation 3.

Table 1: Constants to compute the empirical boundaries of the Habitable Zone using eqn. 1 (see Kopparapu et al.2013).

| Constants | Recent Venus limit Inner edge | Early Mars limit Outer edge |
|---|---|---|
| $S_{sun}$ | 1.7763 | 0.3207 |
| A | 1.4335e-4 | 5.4471e-5 |
| B | 3.3954e-9 | 1.5275e-9 |
| C | -7.6364e-12 | -2.1709e-12 |
| D | -1.1950e-15 | -3.8282e-16 |

$$P = \frac{2\pi \sigma_{SB}^{3/4}}{\sqrt{G}} M^{-1/2} R^{3/2} T_{eff}^3 (S_{Sun} + aT + bT^2 + cT^3 + dT^4)^{-3/4} \quad (3)$$

Models of the HZ take into account that cool red stars emit more light at longer wavelengths and thus heat the surface of a planet more effectively than the same irradiation from a hotter star like the Sun (see Kasting et al. 1993, review: Kaltenegger 2017).

**2.2 Star Selection**

TESS observes sectors of the sky of 24 x 96 degrees. While all stars in the TESS field of view are observed at a 30-minute cadence, a subset of stars were selected for 2-min cadence observations with pixel cutouts (e.g. Rucker et al 2016). The TESS mission developed the TESS Input Catalog (TIC) as a compiled catalog of stellar properties to both select targets for 2-min observations and also to evaluate the resulting observations. The TIC includes about 1.5 billion stars, and a subset of the brightest stars in the TIC was developed into the Candidate Target List (CTL), which includes a more comprehensive set of derived stellar properties. Version 7 of the TIC is described in Stassun et al (2018); the corresponding CTL was used to select target stars for the first year of TESS observations in the southern ecliptic hemisphere. Version 8 of the TIC is described in Stassun et al (2019); the corresponding CTL was used to select target stars for the second year of TESS observations in the northern ecliptic hemisphere. TIC-8 incorporated data from Gaia DR2 (Brown et al. 2018), which improved both distances and thus radii for low-mass stars and differentiating dwarfs from subgiants.

The stars in both CTL-7 and CTL-8 were ranked according to a metric (see e.g. Stassun et al 2019) to identify stars for which TESS could detect small transiting planets. The metric prioritizes bright and small stars, which results in a bimodal distribution of stars in effective temperature, yielding a set of brighter G and K dwarfs, along with another set of fainter M dwarfs. Note that a large fraction of the late K and M dwarfs in the CTL has stellar properties independently determined by Muirhead (2018), which constitutes a subset of the TIC called the Cool Dwarf Catalog.

The list of stars observed by TESS is a combination of targets selected for observation for transit discovery, engineering requirements, asteroseismology, and guest investigator programs. By design, only a small fraction of the millions of stars in the CTL were selected for TESS observations, and some stars were observed by TESS for reasons other than transit detection. We selected the highest-ranked 480,000 stars from CTL-8 and cross-matched them to the list of 232,765 unique stars observed at 2-min cadence in the TESS prime mission, thus identifying 148,420 TESS-observed stars optimally suitable for small-planet transit searches.

For each star in this list, we then calculated the orbital period corresponding to the semimajor axis where a planet would receive similar top of the atmosphere (TOA) irradiation to modern Earth. We also derived the corresponding Earth Analog period ($P_{EA}$) and semimajor axis ($a_{EA}$). Note that this calculation is independent of models of the planet. Stellar masses and radii used were determined in TIC-8, which used Gaia DR2 distances and a set of broadband magnitudes and empirical relations to determine stellar effective temperatures, radii, and masses (see sections 2.3.4 and 2.3.5 of Stassun et al (2019) for the parameter determinations, and section 3.2.2 for the uncertainty determinations).

Then we derive the irradiation at the outer edge of the empirical as defined by Kasting et al (1993) HZ – corresponding to Early Mars (EM) irradiation. The inner edge of the empirical HZ corresponds to Recent Venus's irradiation (RV). The corresponding orbital period ($P_{EM}$ and $P_{RV}$), and semimajor axis ($a_{EM}$ and $P_{EM}$) are derived for an Earth-analog planet in mass, radius, and surface pressure.

Note that we are limiting this analysis to stars with 2-min cadence observations in the TESS prime mission. Thus we are ignored the possibility of detecting planets orbiting in the HZ of stars observed only in the TESS 30-minute full-frame images (FFIs); While there might be some stars in that set suitable for HZ planet detection, the set of targets with 2-min cadence observations have already been preselected for dwarf stars (with some subgiants) ranked according to their suitability for small planet transit detection. Stars observed only in the FFIs are in general fainter, evolved, or earlier spectral type than those in the CTL, making them less amenable for the detection of small planets.

In addition, for all known TOIs (as of Nov 20, 2020) – including host stars that are not part of the Revised TESS HZ catalog – we derive their TOA irradiation the same way as the Earth-analog irradiation described above. We use stellar information from TIC-8 for all TOI hosts to identify any additional TOIs in the HZ.

## 2.3 TESS Observing Constraints

The pattern of overlapping sectors TESS observes means that stars closer to the ecliptic poles are observed longer. In general, when stars are observed in multiple sectors, those sectors are sequential. But that is not always the case. There are small gaps between each TESS CCD and camera, and the probability that a star falls in a gap in a single sector is non-negligible. Thus, a star within 20 degrees of the ecliptic pole could be observed, for instance, over the course of three sectors, with data only in the first and third sectors. Or, a star could be observed both in Sector 1 and Sector 13, with a long gap between.

Furthermore, each individual sector consists of two orbits, with a gap in observations between the orbits of a little over a day in the middle of the sector, along with a gap at the beginning and end of each orbit. These various gaps mean that it is not straightforward to characterize the probability that TESS observed every transit of a long-period planet.

To evaluate the ability to detect periodic transits in the TESS data, we computed several metrics dealing with the TESS observations for each star in the list. For each sector, we assembled the typical number of good observations – those observations not flagged as bad quality by the TESS pipeline – for all stars. Then for each star, we calculated the "Dwell" time as the total accumulated exposure time of good observations. We also calculated the "Duration" time as the time span between the first and last good TESS observations. With these definitions, Duration is always longer than Dwell. The values of Duration and Dwell for each of the prime mission sectors are listed in table 2.

We then also calculated the "Consecutive Sector Duration" (CSD) as the time span between the first and last good observations for the largest number of consecutive sectors in which a star was observed. We adopted CSD as the best way to comprehensively represent the closest quantity to continuous observations by TESS.

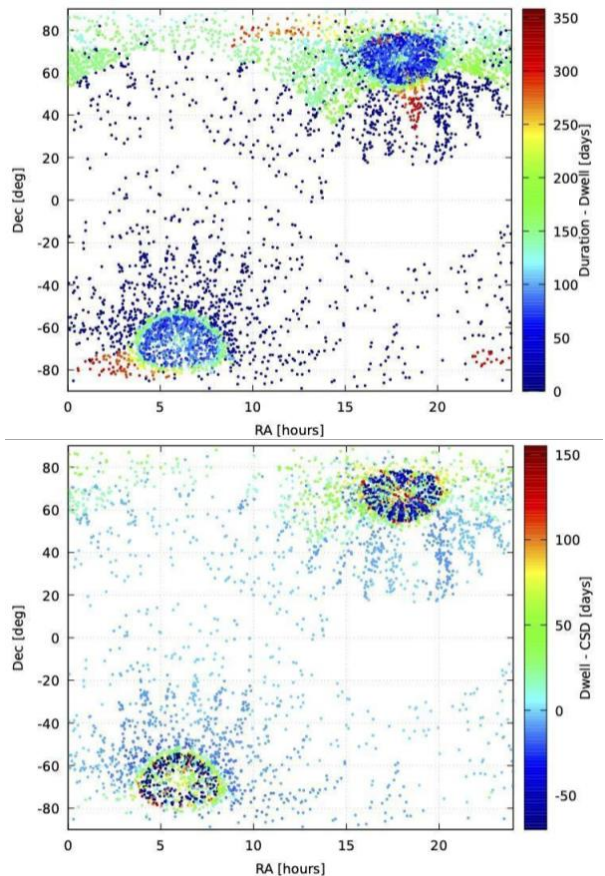

**Figure 1:** TESS observation times for the 4,239 stars in the Revised TESS HZ Catalog listed in table 3: (top) Difference between Duration and Dwell time, (bottom) Difference between Dwell and Continuous Sector Duration (CSD) for all stars shown in equatorial coordinates.

We note again that in no case is a star observed without interruption during this time because there are interruptions between each orbit. But we find that the Dwell time for the largest span of consecutive sectors for a star is usually only about 8% less than CSD. Also, since CSD only considers consecutive sectors, CSD for a star can be much smaller than either Duration or Dwell times.

Figure 1 shows the difference between Duration, Dwell, and CSD. Note that TESS avoided a region of the sky during several sectors in the Northern Hemisphere because the boresight of some of the cameras would have been too close to the Earth and Moon to collect useful data. Stars in this region were not observed during the primary TESS mission. Instead, the spacecraft was pointed farther from the ecliptic, resulting in additional sectors of overlap for some stars nearer to the North Ecliptic Pole.

We first use the Duration to identify the stars that were observed for a minimum of at least 2 $P_{EA}$, which yields 9,823 stars. We then removed stars with bad Gaia data quality flags in TIC-8,[1] yielding 8,977 stars for 2 $P_{EA}$ and 6,492 stars for 3 $P_{EA}$, respectively. We examined the TESS observing patterns for each of the remaining stars in the list of 8,977 stars from the CTL that have good Gaia data quality flags.

After identifying the set of stars where TESS could detect planets with Earth-analog irradiation using the initial cut on Duration, we then apply the more stringent cut in CSD. We find that 4,239 stars were observed with a CSD greater than 3 $P_{EA}$. These represent stars where TESS has the greatest potential to detect planets with Earth-analog irradiation. We find that the mean Duration of a TESS sector is 27.3 days, and the mean Dwell is 25.3 days, and so a rough estimate of the probability that TESS will observe all three transits in such a case is 25.3/27.3 = 93%. More exact calculations can be done on a star-by-star basis but are unnecessary for the overall goals of this paper.

---

[1] The Gaia data quality parameter is described in the beginning of sections 2.3 and 3.1 of Stassun et al. (2019). That parameter is based on the uncertainties in the Gaia photometric and astrometric measurements and is not itself a flag that appears in the Gaia DR2 catalog, but instead is derived from tests introduced in the Gaia mission papers. Specifically, it is described with equations 1 and 2 of the DR2 catalog validation paper (Arenou et al. 2018), which in turn were originally defined in the DR2 HRD paper (Gaia Collaboration 2018), and also in Appendix C of the DR2 astrometric solution paper (Lindegren 2018). After the incorporation of DR2 into the TIC, and the release of TIC-8, that quantity was re-cast in a more limited way in the Gaia dataset, using only the Gaia astrometric errors and referred to as the Renormalized Unit Weight Error (RUWE).

## 3. Results

We provide the list of the 4,329 stars in the Revised TESS Habitable Zone Stellar Catalog in Table 3; figure 2 shows several properties of this set of stars for which TESS can potentially detect planets with Earth-analog irradiation. Because of TESS's observing strategy, the stars are concentrated toward the ecliptic poles where TESS sectors have the greatest mutual overlap.

### 3.1 For how many stars can TESS detect planets receiving Earth-analog irradiation?

TESS has observed 4,329 main sequence stars long enough to see at least three transits of planets receiving the equivalent irradiation to modern Earth. The surface temperatures of these stars range from 2,726 to 4,664 K, with distances between 3 and 207 pc. As expected, the Revised TESS HZ catalog is dominated by cool stars, because a planet receiving Earth-analog irradiation orbits closer to a low luminosity host, resulting in a shorter orbital period; the corresponding orbital periods ($P_{EA}$) around these host stars range between about 4.6 and 119 days.

Using the Dwell metric, we find that the stars are observed for between 18 and 297 days, while their CSD times range from 20.3 to 357 days. Thus, the stars were observed long enough nearly continuously to detect between 3 and 66 transits of planets in orbits receiving the same irradiation as Earth in the TESS data of the primary mission.

The transit depth that would be created by a nominal planet of 1 Earth radius around these stars ranges between about 150 ppm to 11,000 ppm for stars in the Revised TESS HZ catalog (fig.3, left). Because of the short orbital periods of planets around cool stars, even planets of one Earth mass in a circular orbit would produce a radial-velocity semi-amplitude between 0.35 and 63 m s$^{-1}$ (fig.3, right) around these stars, putting them within reach of some current and upcoming spectrographs.

Note that the values we present here for transit depth, as well as radial velocity semi-amplitudes, are a conservative estimate because those increase with planetary radius and mass, respectively. Current estimates for rocky Super-Earths place their maximum radius between 1.6 and 2 Earth radii and the maximum mass at around 18 Earth masses (see e.g., review: Kaltenegger 2017). Planets with larger radii and larger mass will provide larger transit depths and radial velocity semi-amplitudes than calculated here and shown in figure 3.

Table 3 lists the stars sorted by increasing distance and is available online (*filtergraph.com/tesshzstars*). It provides the stellar properties as well as orbital distances ($a_{EA}$) and orbital periods ($P_{EA}$) of a nominal planet receiving similar irradiation to modern Earth.

### 3.2 For how many stars can TESS detect planets out to Mars' orbit – the full Habitable Zone?

In addition to observing planets with Earth-equivalent irradiation, TESS can explore the population of transiting planets throughout the HZ.

For a subset of 614 stars, TESS could detect up to 3 transits for planets throughout the full empirical HZ, out to the equivalent orbit of a young Mars. The surface temperatures of this subset of stars are cooler than the full main catalog: 2,780 K to 3,766 K with distances between 7 and 130 pc, observed for between 63 and 297 days (Dwell). If the orbital period of detected planet will be larger than $P_{RV}$ and smaller than $P_{EM}$, the detected planet will orbit its host star in the star's empirical HZ.

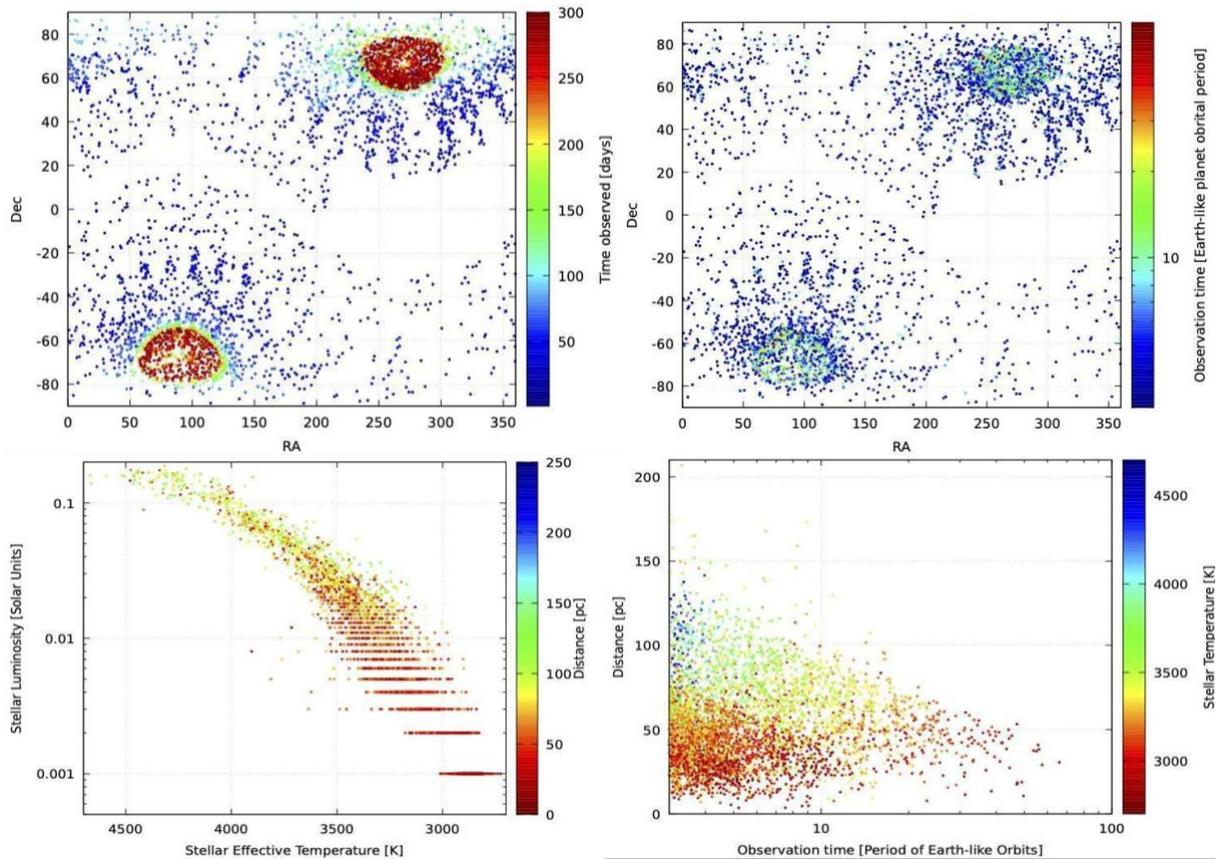

**Figure 2**: Properties of the 4,239 stars in the Revised TESS HZ catalog, listed in Table 3, which were observed for a minimum of 3 $P_{EA}$: (top left) Dwell time for stars shown in equatorial coordinates. (top right) Dwell time scaled to units of Earth-**analog** orbital period ($P_{EA}$). (bottom left) HR Diagram, with the distance shown on the color axis. (bottom right) Distance versus Dwell time in scaled units of $P_{EA}$, with stellar temperature on the color axis. Note that the horizontal structures at the end of the main sequence are artifacts of the calculations of stellar properties for cool stars used for the Cool Dwarf Catalog, discussed in Muirhead et al (2018), and imported to TIC-8.

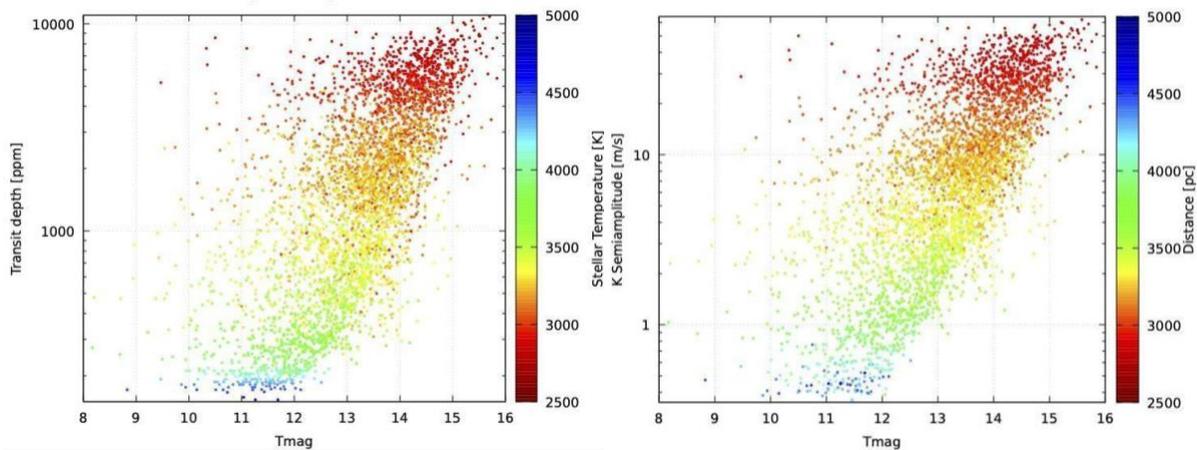

**Figure 3**: Detection metrics for the 4,239 stars in the Revised TESS HZ catalog, listed in Table 3 and shown in figure 2 – assuming the planets are Earth analogs (1 $M_E$, 1 $R_E$, and $P_{EA}$): (left) Transit depths and (right) Radial Velocity semi-amplitudes, in both cases plotted versus the host star TESS magnitude. Distance is given on the color bar.

All stars were observed long enough continuously (CSD) to observe between a minimum of 3 and 17 transits of planets throughout the HZ in the TESS data of the primary mission. Table 3 provides the stellar properties as well as orbital distances ($a_{EM}$) and orbital periods ($P_{EM}$) of a nominal planet at the outer edge of the empirical HZ for all 614 stars.

The transit depth of a nominal Earth-size planet for this subset of stars in the Revised TESS HZ Catalog ranges between about 330 and 10,200 ppm. The minimum radial-velocity semi-amplitude of a nominal Earth-mass planet on the outer edge of the HZ ($a_{EM}$) ranges between 0.56 and 15 m s$^{-1}$. These values increase with planetary radius and mass, respectively, as discussed earlier.

### 3.3 Known Host Stars and TOIs

Of the 4,239 stars in the Revised TESS HZ catalog, nine host known planets – GJ 1061, GJ 1132, GJ 3512, GJ 685, Kepler-42, LHS 1815, L_98-59, RR Cae, TOI 700 (see Table 4) – around which TESS could identify additional companions receiving Earth-**analog** irradiation. GJ 1061c and GJ 1061d at 3.7 pc (Anglada-Escudé et al. 2020) as well as TOI 700d at 40 pc (Gilbert et al. 2020) orbit their star in the empirical HZ. Additionally, 37 of the stars in Table 4 are currently (as of Nov 20, 2020) listed as unconfirmed TOIs, located between 10.6 and 115 pc. Properties of known host stars and TOIs in the Revised TESS HZ Catalog are given in Table 4.

Three of these TOIs orbit in the HZ of their star – TOIs 203.01, TOI 715.01, and TOI 2298.01 (see Table 4). TOI 203.01, at 24.8pc, orbits its host star with an orbital period of 52 days and receives about 0.21 times Earth's irradiation. It has a radius of 1.2 Earth radii. TOI 715.01 at 42.4 pc, orbits its host star in 19.3 days, and receives about 0.74 times Earth's irradiation, and has a radius of about 1.6 Earth radii. TOI 2298.01 orbits its host star in about 165.2 days and receives about 0.48 times Earth's irradiation. It is about 3 times Earth's radius and thus less likely to be a rocky planet but could potentially harbor a rocky moon (see e.g. Williams et al. 1997, Kaltenegger 2010, Barnes & Heller 2013, Kipping et al 2014, Martinez-Rodriguez et al 2019).

We have reviewed the properties of the three TOIs we identified in the HZ of their stars. TOI 715.01 appears to be a good candidate. We found that while TOI 203.01 would nominally be located in the HZ if the listed properties of the TOI are accurate, it is likely a false alarm. Three of the four supposed transit events do not look real; the first and third events coincide with pointing jumps caused by planned spacecraft thruster fires, and the fourth is not present. Only the second event observed by TESS appears to be real, but if it is, the period would be unknown and would be unlikely to correspond to the HZ. TOI 2298.01 suffers from a similar issue, as TESS appears to have observed only a single transit. However, the duration of that transit is consistent with a period in the HZ, albeit with very large uncertainty. We recommend further investigation of both TOI systems.

As stated in Section 2.2, we have re-derived the incident flux for all TOIs using updated stellar parameters from TIC-8. Using that information, we find that five additional TOIs nominally orbit in the HZ of their host stars: TOIs 256.01, 793.01, 798.01, 1227.01, and 1823.01. However, their host stars are not included in the Revised TESS HZ star catalog. We investigate why:

TOI 256.01 is a known transiting planet (LHS 1140 b; Dittman et al. 2017) with an orbital period of 24.7 days. While the known transit signal was seen in the TESS light curve and marked as a TOI, the star was only observed with a Duration of 20.26 d, and so was not observed by TESS long enough to see three transits.

TOI 793.01 is listed with a period of 164 d. It has a Duration of 357 d, and so was not observed long enough to measure three transits. But upon inspection of the TESS light curve, it appears that the true period of the TOI signal is 54 days, thus the planet candidate would be too hot to be in the HZ.

TOI 798.01 is not in the CTL. The star does not have a Gaia DR2 parallax and was selected for 2-minute TESS observations via Guest Investigator proposal G011180 (C. Dressing). We note that the lack of a five-parameter astrometric solution could indicate the presence of another star in the system, which comes with an elevated risk of astrophysical false positives or modifications to the planet properties, especially if the transiting object orbits a different star. TOI 1227.01 has a low priority in the CTL, due to the faintness of the host star (T=13.8). Thus, it is below the cutoff threshold in CTL priority we imposed described in section 2.2. It resides in a crowded field and was selected for TESS observations via Guest Investigator proposal G011180.

TOI 1823.01 has a listed orbital period of 194 d. The TESS observation Duration is 110.1 days. We discuss this case in more detail in Section 4.2.

With further analysis of the TESS data, we expect the number of TOIs as well as the number of confirmed planets in the HZ to increase. The Revised TESS HZ Catalog (Table 3) provides a tool to identify stars around which TESS could detect potentially habitable worlds.

# 4. Discussion

We identify the 4,239 stars for which the TESS mission has collected enough data in its primary mission to detect planets with Earth-equivalent irradiation if they transit, and if the transits have high enough SNR. For a subset of 614 of those stars, TESS has collected enough data to observe planets out to a Mars' analog orbit, thus covering the full empirical Habitable Zone of those systems.

## 4.1 Missed a transit?

Note that we assume here that the required observation time is three times the orbital period of a planet receiving the same TOA irradiation as modern Earth (CSD > 3 $P_{EA}$). For some stars, one of the three transits could have fallen into the observing gaps between orbits.

How does the number of stars change if we increase the requirements to four transits – to account for one of the transits not being observed? The number of stars that TESS has observed continuously for a minimum of 4 $P_{EA}$ decreases from 4,239 to 2,742 stars. The number of stars TESS has observed for a minimum of 4 $P_{EM}$ decreases to 339 stars. Note that the likelihood of seeing four transits increases linearly with CSD: e.g. 10% of systems observed for 3.1 $P_{EA}$ will show four transits. However, two detectable transits should call attention to a planet in an Earth-analog orbit and follow-up observations from the ground could detect the third transit. Also, the TESS extended mission will provide an opportunity to revisit parts of the sky and observe some systems again.

If we relax the requirements to only two transits to call attention to the system, then the number of stars increases to 7,297 stars for the minimum CSD of 2 $P_{EA}$ and 1,082 stars for a minimum CSD of 2 $P_{EM}$.

## 4.2 Additional stars which have not been observed continuously

The observation time (Duration) of 6,492 stars is longer than three times the orbital period of a planet receiving similar irradiation from its star to modern Earth ($P_{EA}$). However, for 2,253 stars, these observations, while longer than 3 $P_{EA}$, were not continuous, which means they were not observed in consecutive sectors (CSD < 3 $P_{EA}$). The spacing of these observations compared to the orbital period of a planet with Earth-analog irradiation requires attention to the individual stars to assess how many such transits could have been detected – which depends strongly on the assumption when the transit could occur. We have not included these stars in the Revised TESS Habitable Zone Catalog, but the list can be found in Table 5 (available online *filtergraph.com/tesshzstars*). Of the current TOI list, TOI 1823.01 (listed above in section 3.3) represents one such case. For that candidate, the two observed transit events suggest a planet in the HZ of the host star, although gaps in the TESS observations leave open the possibility of a shorter-period planet well inside the HZ. We recommend further investigation of this TOI 1823.01.

### 4.3 The influence of the uncertainty of stellar parameters on our results

Several teams have shown the dependence of the HZ boundaries and any planet's position in the HZ on the uncertainty of the stellar properties (e.g., Kaltenegger & Sasselov 2011, Kane et al 2014). Following Kane et al (2014) we derive the uncertainties on the calculated orbital periods by propagating the errors on stellar mass, radius, and effective temperature. We find that the typical uncertainty on the Earth-analog as well as the HZ boundary periods are about 15% with a maximum of 30% and a minimum of 11%.

We have included the stellar parameter uncertainties as well as the uncertainty of the calculated periods for all 8,977 stars TESS has observed - continuously as well as not continuously - for a minimum of three times the Earth-analog period (Duration > 3 $P_{EA}$) in table 6.

### 4.4 Differences with the initial TESS HZ catalog

Our current analysis supersedes the initial catalog for two reasons. First, it is based on the actual observation time per star (see Table 2), which turned out to be very different for some stars from what was predicted using the earlier, simplified calculation for the length of TESS observations, especially in sectors where stray-light issues changed the TESS observing strategy, as noted in section 2.3. The effect on observation time of the exact position of stars in overlapping sectors or falling into CCD gaps is accounted for in the new catalog.

Second, this analysis uses the full incorporation of the Gaia DR2 catalog into the TIC, which had not yet been completely incorporated into the version of the TESS Input Catalog (TIC-7.2) in in previous paper, Kaltenegger et al (2019). With TIC-8 using all of Gaia DR2, especially the DR2 quality flags, the Revised TESS HZ catalog now provides more reliable stellar parameters.

In the prior paper, we calculated an estimated SNR per transit for each of the stars in the catalog. Those estimates were based on a TESS photometric noise model. With the actual TESS data available, we decided not to include a SNR estimate per transit in this paper. Depending on the data pipeline and detrending tools that investigators wish to use, people can calculate the real noise properties in the photometry for each star of interest, accounting for both random and systematic sources of noise. Such a broad analysis is outside the scope of this paper, but we recommend that anyone who plans to do so consults the TESS data release notes.[2]

### 4.5. Prospects for follow-up characterization of TESS planets

Because of the short periods of planets receiving similar irradiation to Earth from a cool host star, even an Earth-mass planet produces a radial-velocity semi-amplitude between 0.35 and 63 m s$^{-1}$ (see Fig.3), with more massive rocky super-Earths producing greater values. For a nominal 10 $M_E$ super-Earth these values increase by a factor of 10 to 3.5 to 630 m s$^{-1}$. The transit depth ranges between 150 and 11,000 ppm for a nominal Earth-size planet in the Revised TESS HZ

---
[2] https://archive.stsci.edu/tess/tess_drn.html

Catalog (see Fig.3) and increases to 240 to 17,600 for a super-Earth of 1.6 Earth radii. Such measurements are within reach of current and upcoming spectrographs (see discussion in Kaltenegger et al. 2019)

Upcoming ground- and space-based facilities can further improve on measurements of stellar and planetary parameters – e.g. with CHEOPS (Fortier et al. 2014) and PLATO (Rauer et al. 2014). While observations of atmospheric composition for Earth-like planets with upcoming ground-based Extremely Large telescopes and space-based telescopes like JWST will be challenging, they are for the first time possible (e.g. Clampin et al 2009, Deming et al 2009, Kaltenegger & Traub 2009, Rauer et al. 2011, Snellen et al 2013, 2015, Rodler & Lopez-Morales 2014, Kempton et al. 2018, Batalha et al 2018, Lopez-Morales et.al.2019, Serindag & Snellen 2019, Lin & Kaltenegger 2020).

Because TESS focuses on close-by stars, the apparent angular size of a planetary orbit which receives Earth irradiation around the stars in the Revised TESS Habitable Zone Catalog ranges between 0.6 mas and 17.3 mas. A telescope with an inner working angle (IWA) of 6 mas – e.g. the 38m-diameter ELT (assuming $\theta_{IWA} \approx 2(\lambda/D)$, where $\lambda$ is the observing wavelength and D is the telescope diameter) – could resolve a planet receiving Earth-analog irradiation around 90 of the 4,239 stars.

## 5. Conclusions

The Revised TESS Habitable Zone Catalog contains 4,239 stars that have been observed nearly continuously for at least three times the orbital period where a planet would receive Earth-equivalent stellar irradiation. If any of these stars host such a planet, the transit signal could be detectable in the TESS photometry, and in some cases, the planet could be dynamically detected with current or near-term RV facilities. The angular separation associated with such an orbit could be resolved by upcoming telescopes like the ELT for about 90 of the stars in this catalog.

For a subset of 614 out of the 4,239 stars, TESS has observed the star long enough to be able to observe planets throughout the full temperate, empirical habitable zone out to the equivalent of Mars' orbit.

Of the 4,239 stars in the Revised TESS HZ catalog, TESS has already discovered and confirmed one small planet in the habitable zone of its star (TOI 700d). This is one of 9 known exoplanet hosts in the Revised TESS HZ Catalog around which TESS could identify additional companions with Earth-analog irradiation. 37 additional stars host yet unconfirmed TOIs, with at least one of these candidates (TOI 715) orbiting in the HZ of its star.

We created this catalog as a useful tool that will allow scientists to prioritize TESS stars for intensive analysis and follow-up observations. They represent the best opportunities to find potentially Earth-like planets in the TESS data if they exist – on our path to find life in the cosmos.

**Acknowledgment:**
LK acknowledges support from the Carl Sagan Institute at Cornell and the Brinson Foundation. This paper includes data prepared for the TESS mission. Funding for the TESS mission is provided by the NASA Explorer Program. This work uses results from the European Space Agency (ESA) space mission Gaia. Gaia data are being processed by the Gaia Data Processing and Analysis Consortium (DPAC). Funding for the DPAC is provided by national institutions, in particular the institutions participating in the Gaia MultiLateral Agreement (MLA). The Gaia mission website is https://www.cosmos.esa.int/gaia. The Gaia archive website is https://archives.esac.esa.int/gaia

# References

Anglada-Escudé, G.; Reiners, A.; Pallé, E.; Ribas, I.; Berdiñas, Z. M. et al, MNRAS, 493 (1), 536 (2020)

Arenou F., et al., 2018, Gaia Data Release 2. Catalog validation, A&A, 616, A17 (2018)

Barnes, R., & Heller, R., Astrobiology, 13, 279 (2013)

Berger, T. A., Huber, D., Gaidos, E., & van Saders, J. L., 2018, ApJ, 866, 99 (2018)

Clampin M., JWST Science Working Group, JWST Transits Working Group, D. Deming, and D. Lindler, Comparative Planetology: Transiting Exoplanet Science with JWST, AST2010 Science Whitepaper JWST Transit Science. (2009)

Deming D., S. Seager, J. Winn, et al., Publ Astron Soc Pac 121:952–967 (2009)

Dittmann, J.A, Irwin, J.M. Charbonneau, D. et al. *Nature* volume 544, pages333–336 (2017)

Fischer, D. A., Anglada-Escude, G., Arriagada, P. et al., PASP, 128, 066001(2016)

Fortier, A., Beck, T., Benz, W., et al., Proc. SPIE, 9143, 91432J (2014)

Gaia Collaboration; Brown, A. G. A.; Vallenari, A., et al.; A&A, 616, A1, 22 (2018)

Gilbert, E.A., Barclay, T., Schlieder, J.E., et al. AJ,160, 3 (2020)

Guerrero, N.M, Seager, S., Huang, C.S., et al. AJ, submitted

Huang, C.X., Burt, J., Vanderburg, et al., ApJL, 868, 2, L39 (2018)

Johns, D., Marti, C., Huff, M., et al., ApJS, 239, 14 (2018)

Kaltenegger L, Traub WA. *Ap. J.* 698:519 (2009)

Kaltenegger, L. ApJL, 712, L125 (2010)

Kaltenegger, L., Annual Review of Astronomy and Astrophysics 55, 433-485, (2017)

Kaltenegger, L., Pepper, J., Stassun, K., Oelkers, R., ApJL 874 (1), L8, (2019)

Kaltenegger, L.& Haghighipour N. ApJ, 777, 2, 165 (2013)

Kaltenegger, L.& Sasselov, D., ApJL, 736, 2, L25, 6 (2011)

Kane, S. R., & Hinkel, N. R., ApJ, 762, 7 (2013)

Kane, S. R., ApJ, 782, 2, 111, 8 (2014)

Kasting, J.F., Whitmire, D.P. & Reynolds, R.T., Icarus, 101:108–28 (1993)

Kempton E., et al, Publications of the Astronomical Society of the Pacific, 130, 993, 114401 (2018)

Kipping, D., Nesvorný, D., Buchhave, L. A., Hartman, J., Bakos, G. Á. & Schmitt, A. R, ApJ, 784, 28 (2014)

Kopparapu, R. K., Ramirez, R. M., Schottel Kotte, J., Kasting, J. F., Domagal-Goldman, S., & Eymet, V., ApJ, 787,2, L29 (2014)

Kopparapu, R. K., Ramirez, R., Kasting, J. F., et al., ApJ, 765, 2, 131. (2013)

Kostov, V.B, Schlieder, J.E., Barclay, T., et al, AJ, 158, 1, 32 (2019)

Lin. Z. & Kaltenegger, L., High-resolution reflection spectral models of Proxima-b and Trappist-1e, MNRAS, 491 (2), 2845-2854 (2020)

Lindegren, L. et al, 2018, Gaia Data Release 2. The astrometric solution, A&A 616, A2 (2018)

Lopez-Morales, M., Currie, T., Teske, J., et al., BAAS, 51, 162 (2019)

Luque, R., Pallé, E., Kossakowski, D., et al., A&A, 628, A39 (2019)

Martinez-Rodriguez et al., ApJ, 887, 261 (2019)

Muirhead, P.S., et al, A Catalog of Cool Dwarf Targets for the Transiting Exoplanet Survey Satellite, AJ, 155, 4, 180, 14 (2018)

Pierrehumbert, R., & Gaidos, E.,ApJL,734, L13 (2011)

Ramirez RM & Kaltenegger L.. ApJ. 823(1):6, POST-MS (2016)

Ramirez, R.M.& Kaltenegger, L., ApJL 837 (1), L4, 60 (2017)

Ramirez, R.M.& Kaltenegger, L., ApJL 858 (2), 72 (2018)

Ramirez, R.M.& Kaltenegger, L., ApJL, 797, 2, article id. L25, 8, (2014)

Rauer H, Gebauer S, Paris P V, Cabrera J, Godolt M, et al. Astron. Astrophys. 529(5):A8 (2011)

Rauer, H., Catala, C., Aerts, C., et al, Experimental Astronomy, 38, 249 (2014)

Ricker, G. R., et al. Proceedings of the SPIE, Volume 9904, id. 99042B 18 pp. (2016)

Rodler, F., & Lopez-Morales, M., ApJ, 781, 54 (2014)

Rodriguez, J.E., Vanderburg, A., Zieba, S., Kreidberg, L. et al., AJ, 160, 3, (2019)

Serindag, D.B. & Snellen, I.A.G., ApJL, 871, 1, L7, 5, (2019)

Snellen, I., de Kok, R., Le Poole, R., Brogi, M., & Birkby, J., ApJ, 764, 182 (2013)

Snellen, I., de Kok, R., Birkby, J. L., et al., A&A, 576, A59 (2015)

Stassun K., et al., TIC-8, https://arxiv.org/abs/1905.10694 (2019)

Stassun, K. G.; Oelkers, R. J.; Pepper, J. et al. AJ, 156, 3, article id. 102, 39 pp. (2018).

Stevenson, D. J., Natur, 400, 32 (1999)

Vanderspek, R., Huang, C.X.,Vanderburg, A. et al, ApJL 871, 2, L24 (2019)

Williams, D.M., Kasting J.F & Wade, R.A. *Nature* 385, 234 (1997)

**Table 2:** Metrics for TESS observation time per sector in days

| #Sec | Duration | Dwell | #Sec | Duration | Dwell | #Sec | Duration | Dwell | #Sec | Duration | Dwell |
|---|---|---|---|---|---|---|---|---|---|---|---|
| 1 | 27.88 | 25.17 | 8 | 24.64 | 18.64 | 15 | 26.04 | 24.79 | 22 | 26.88 | 24.75 |
| 2 | 27.40 | 25.44 | 9 | 24.47 | 22.72 | 16 | 24.67 | 23.35 | 23 | 25.55 | 21.50 |
| 3 | 20.27 | 18.68 | 10 | 24.80 | 21.79 | 17 | 23.79 | 21.55 | 24 | 26.49 | 22.51 |
| 4 | 25.94 | 19.98 | 11 | 25.91 | 23.34 | 18 | 23.84 | 21.60 | 25 | 25.67 | 23.95 |
| 5 | 25.59 | 22.78 | 12 | 25.26 | 21.84 | 19 | 25.06 | 23.45 | 26 | 24.87 | 23.53 |
| 6 | 21.77 | 20.60 | 13 | 25.91 | 24.69 | 20 | 26.32 | 22.97 | | | |
| 7 | 24.45 | 22.73 | 14 | 26.85 | 18.84 | 21 | 27.35 | 24.11 | | | |

**Table 3**: The Revised TESS HZ Catalog: 4,239 stars TESS observed long enough to see **planets with Earth-analog irradiation**

| Units | Label | Explanations |
|---|---|---|
| --- | TIC | TESS Input Catalog identifier |
| K | Teff | Effective temperature |
| mag | Tmag | TESS broad band magnitude |
| pc | Dis | Distance |
| d | Obs | Observing time |
| d | Obs | Real observing time per star (Dwell) |
| d | Obs | Consecutive Duration Observing time (CSD) |
| AU | aEA | Earth Analog irradiation orbital separation |
| AU | aEM | Early Mars irradiation orbital separation |
| AU | aRV | Recent Venus irradiation orbital separation |
| d | PerEA | Earth Analog irradiation orbital period |
| d | PerEM | Early Mars irradiation orbital period |
| d | PerRV | Recent Venus irradiation orbital period |
| deg | ELAT | Ecliptic latitude |
| deg | GLAT | Galactic latitude |
| deg | RAdeg | Right Ascension in decimal degrees (J2000) |
| deg | DEdeg | Declination in decimal degrees (J2000) |
| h | CSD | Consecutive Sector duration observing time |
| h | Dwell | Dwell observing time |
| h | Duration | Duration observing time |
| --- | Gaia | Gaia catalog identifier |
| --- | 2MASS | 2MASS all sky survey catalog identifier |

data available online and on *filtergraph.com/tesshzstars*

**Table 4**: Known planets and TOIs in the Revised TESS HZ Catalog

| TIC_ID | Host Name | d [pc] | Teff [K] | Name | Period [days] | PerEA [days] | HZ |
|---|---|---|---|---|---|---|---|
| 63126862 | Kepler-42 | 40.06 | 3357 | Kepler-42 b | 1.2 | 12.7 | 0 |
| 63126862 | Kepler-42 | 40.06 | 3357 | Kepler-42 c | 0.5 | 12.7 | 0 |
| 63126862 | Kepler-42 | 40.06 | 3357 | Kepler-42 d | 1.9 | 12.7 | 0 |
| 219780306 | GJ_685 | 14.32 | 3753 | GJ 685 b | 24.2 | 56.3 | 0 |
| 445064836 | GJ_3512 | 9.49 | 2902 | GJ 3512 b | 203.6 | 7.2 | 0 |
| 79611981 | GJ_1061 | 3.67 | 2905 | GJ 1061 b | 3.2 | 7.3 | 0 |
| 79611981 | GJ_1061 | 3.67 | 2905 | GJ 1061 c | 6.7 | 7.3 | 1 |
| 79611981 | GJ_1061 | 3.67 | 2905 | GJ 1061 d | 13.0 | 7.3 | 1 |
| 101955023 | GJ_1132 (TOI 667) | 12.61 | 3261 | GJ 1132 b | 1.6 | 14.9 | 0 |
| 101955023 | GJ_1132 (TOI 667) | 12.61 | 3261 | GJ 1132 c | 8.9 | 14.9 | 0 |
| 260004324 | LHS_1815 (TOI 704) | 29.84 | 3596 | LHS 1815 b | 3.8 | 44.5 | 0 |
| 307210830 | L_98-59 (TOI 175) | 10.62 | 3429 | L 98-59 b | 2.3 | 24.1 | 0 |
| 307210830 | L_98-59 (TOI 175) | 10.62 | 3429 | L 98-59 c | 3.7 | 24.1 | 0 |
| 307210830 | L_98-59 (TOI 175) | 10.62 | 3429 | L 98-59 d | 7.5 | 24.1 | 0 |
| 219244444 | RR_Cae | 21.19 | 3904 | RR Cae b | 4343.5 | 21.8 | 0 |
| 150428135 | TOI 700 | 31.13 | 3494 | TOI 700 b | 10.0 | 34.0 | 0 |
| 150428135 | TOI 700 | 31.13 | 3494 | TOI 700 c | 16.1 | 34.0 | 0 |
| 150428135 | TOI 700 | 31.13 | 3494 | TOI 700 d | 37.4 | 34.0 | 1 |
| 307210830 | TOI 175 | 10.62 | 3429 | TOI 175.01 | 3.7 | 24.06 | 0 |
| 307210830 | TOI 175 | 10.62 | 3429 | TOI 175.02 | 7.5 | 24.06 | 0 |
| 307210830 | TOI 175 | 10.62 | 3429 | TOI 175.03 | 2.3 | 24.06 | 0 |
| 259962054 | TOI 203 | 24.79 | 3209 | TOI 203.01 | 52.0 | 14.6 | 1 |
| 55650590 | TOI 206 | 47.75 | 3383 | TOI 206.01 | 0.7 | 25.87 | 0 |
| 141608198 | TOI 210 | 42.81 | 3264 | TOI 210.01 | 9.0 | 21.81 | 0 |
| 32090583 | TOI 218 | 52.34 | 3249 | TOI 218.01 | 0.4 | 18.54 | 0 |
| 220479565 | TOI 269 | 57.02 | 3546 | TOI 269.01 | 3.7 | 33.71 | 0 |
| 260708537 | TOI 486 | 15.21 | 3470 | TOI 486.01 | 1.7 | 33.87 | 0 |
| 200322593 | TOI 540 | 14.00 | 3130 | TOI 540.01 | 1.2 | 11.48 | 0 |
| 101955023 | TOI 667 | 12.61 | 3261 | TOI 667.01 | 1.6 | 14.92 | 0 |
| 141527579 | TOI 698 | 63.37 | 3526 | TOI 698.01 | 15.1 | 39.3 | 0 |
| 260004324 | TOI 704 | 29.84 | 3596 | TOI 704.01 | 3.8 | 44.53 | 0 |
| 271971130 | TOI 715 | 42.40 | 3187 | TOI 715.01 | 19.3 | 14.13 | 1 |
| 277634430 | TOI 771 | 25.28 | 3306 | TOI 771.01 | 2.3 | 16.6 | 0 |
| 300710077 | TOI 789 | 43.41 | 3461 | TOI 789.01 | 5.4 | 29.23 | 0 |
| 271596225 | TOI 797 | 56.19 | 3607 | TOI 797.01 | 1.8 | 42.23 | 0 |
| 271596225 | TOI 797 | 56.19 | 3607 | TOI 797.02 | 4.1 | 42.23 | 0 |
| 38460940 | TOI 805 | 51.88 | 3366 | TOI 805.01 | 4.1 | 20.37 | 0 |
| 198212955 | TOI 1242 | 110.02 | 4255 | TOI 1242.01 | 0.4 | 104.15 | 0 |
| 229781583 | TOI 1245 | 80.1 | 3692 | TOI 1245.01 | 4.8 | 50.15 | 0 |
| 235683377 | TOI 1442 | 41.17 | 3328 | TOI 1442.01 | 0.4 | 21.62 | 0 |
| 377293776 | TOI 1450 | 22.45 | 3407 | TOI 1450.01 | 2.0 | 35.58 | 0 |
| 420112589 | TOI 1452 | 30.52 | 3248 | TOI 1452.01 | 11.1 | 17.92 | 0 |
| 165551882 | TOI 1633 | 48.05 | 3429 | TOI 1633.01 | 12.2 | 24.76 | 0 |
| 259168516 | TOI 1680 | 37.22 | 3231 | TOI 1680.01 | 4.8 | 13.68 | 0 |
| 1884091865 | TOI 1697 | 83.74 | 4067 | TOI 1697.01 | 10.7 | 94.54 | 0 |
| 198206613 | TOI 1741 | 61.93 | 4324 | TOI 1741.01 | 10.9 | 94 | 0 |
| 219860288 | TOI 1743 | 41.28 | 3281 | TOI 1743.01 | 4.3 | 21.48 | 0 |

| TIC_ID | Host Name | d [pc] | Teff [K] | Name | Period [days] | PerEA [days] | HZ |
|---|---|---|---|---|---|---|---|
| 233602827 | TOI 1749 | 99.56 | 3959 | TOI 1749.01 | 4.5 | 66.18 | 0 |
| 233602827 | TOI 1749 | 99.56 | 3959 | TOI 1749.02 | 9.0 | 66.18 | 0 |
| 287139872 | TOI 1752 | 30.52 | 3652 | TOI 1752.01 | 0.9 | 48.76 | 0 |
| 332477926 | TOI 1754 | 81.55 | 3849 | TOI 1754.01 | 16.2 | 64.27 | 0 |
| 441739871 | TOI 1763 | 88.07 | 3520 | TOI 1763.01 | 3.8 | 39.18 | 0 |
| 1551345500 | TOI 1764 | 87.29 | 4415 | TOI 1764.01 | 47.4 | 100.65 | 0 |
| 198385543 | TOI 1846 | 47.25 | 3512 | TOI 1846.01 | 3.9 | 33.86 | 0 |
| 441738827 | TOI 2084 | 114.55 | 3630 | TOI 2084.01 | 6.1 | 42.42 | 0 |
| 441738827 | TOI 2084 | 114.55 | 3630 | TOI 2084.02 | 8.1 | 42.42 | 0 |
| 356016119 | TOI 2094 | 50.02 | 3457 | TOI 2094.01 | 18.8 | 29.57 | 0 |
| 235678745 | TOI 2095 | 41.92 | 3746 | TOI 2095.01 | 17.7 | 46.26 | 0 |
| 235678745 | TOI 2095 | 41.92 | 3746 | TOI 2095.02 | 28.2 | 46.26 | 0 |
| 142748283 | TOI 2096 | 48.48 | 3285 | TOI 2096.01 | 3.1 | 16.66 | 0 |
| 142748283 | TOI 2096 | 48.48 | 3285 | TOI 2096.02 | 6.4 | 16.66 | 0 |
| 389900760 | TOI 2120 | 32.18 | 3179 | TOI 2120.01 | 5.8 | 14.66 | 0 |
| 229608594 | TOI 2298 | 84.17 | 4239 | TOI 2298.01 | 165.2 | 95.50 | 1 |

| Units | Label | Explanations |
|---|---|---|
| --- | TIC_ID | TESS Input Catalog identifier |
| --- | Host Name | star name |
| K | Teff | Stellar effective temperature |
| --- | TOI | Name of planet or TOI |
| d | Period | orbital period |
| d | PerEA | Earth Analog irradiation orbital period |
| --- | HZ | planet or TOI in the HZ (1) |

Note (1):
   1 = yes;   0 = no.
--------------------------------------------------------------------------------
data available online and on *filtergraph.com/tesshzstars*

**Table 5**: 2,253 Stars TESS observed long enough – but not in consecutive sectors – to see planets with Earth-analog irradiation

| Units | Label | Explanations |
|---|---|---|
| --- | TIC_ID | TESS Input Catalog identifier |
| K | Teff | Effective temperature |
| mag | Tmag | TESS broad band magnitude |
| pc | Dis | Distance |
| d | Obs | Observing time |
| d | Obs | Real observing time per star (Dwell) |
| d | Obs | Consecutive Duration Observing time (CSD) |
| AU | aEA | Earth Analog irradiation orbital separation |
| AU | aEM | Early Mars irradiation orbital separation |
| AU | aRV | Recent Venus irradiation orbital separation |
| d | PerEA | Earth Analog irradiation orbital period |
| d | PerEM | Early Mars irradiation orbital period |
| d | PerRV | Recent Venus irradiation orbital period |
| deg | ELAT | Ecliptic latitude |
| deg | GLAT | Galactic latitude |
| deg | RAdeg | Right Ascension in decimal degrees (J2000) |
| deg | DEdeg | Declination in decimal degrees (J2000) |
| h | CSD | Consecutive Sector observing time |
| h | Dwell | Dwell observing time |
| h | Duration | Duration observing time |
| --- | Gaia | Gaia catalog identifier |
| --- | 2MASS | 2MASS all sky survey catalog identifier |

data available online and on *filtergraph.com/tesshzstars*

**Table 6:** Uncertainties in stellar parameters and calculated periods for all 8,977 Stars TESS observed longer than three times the Earth-analog orbital period, continuous and non-continuous.

| Units | Label | Explanations |
|---|---|---|
| --- | TIC | TESS Input Catalog identifier |
| --- | fr_lum_err | fractional error in stellar luminosity |
| --- | fr_mass_err | fractional error in stellar mass |
| --- | fr_radius_err | fractional error in stellar radius |
| --- | fr_P_EA_err | fractional error of the Earth-analog period |
| --- | fr_P_RV_err | fractional error of the Recent Venus period |
| --- | fr_P_EM_err | fractional error of the Early Mars period |

data available online and on *filtergraph.com/tesshzstars*